\documentclass{ws-procs9x6}
\usepackage{amssymb,amsmath}
\usepackage{graphicx}

\def\gev{{\rm \, Ge\kern-0.125em V}}
\def\tev{{\rm \, Te\kern-0.125em V}}
\def\beq{\begin{equation}}
\def\eeq{\end{equation}}
\def\ba{\begin{eqnarray}}
\def\ea{\end{eqnarray}}

\begin{document}


\title{Probing  Patterns of Supersymmetry Breaking using 
Phenomenological Constraints\footnote{\uppercase{T}alk presented 
at {\it \uppercase{SUSY} 2003:
\uppercase{S}upersymmetry in the \uppercase{D}esert}\/, 
held at the \uppercase{U}niversity of \uppercase{A}rizona,
\uppercase{T}ucson, \uppercase{AZ}, \uppercase{J}une 5-10, 2003.
\uppercase{T}o appear in the \uppercase{P}roceedings.}}

\author{Vassilis~C.~Spanos}

\address{William I. Fine Theoretical Physics Institute, \\
University of Minnesota, Minneapolis, MN 55455, USA}


\maketitle

\abstracts{Specific models of supersymmetry breaking predict 
relations between the
trilinear and bilinear soft supersymmetry breaking parameters $A_0$ and
$B_0$ at the input scale.
Models with $A_0 = B_0 + m_0$ as well as the  simplest
Polonyi model with $A_0 = (3 - \sqrt{3})m_0 $ are discussed.
In such cases, the value of $\tan \beta$ can be
calculated as a function of the scalar masses $m_0$ and the gaugino masses
$m_{1/2}$, and  various experimental constraints
 can be applied to constrain it.}


One of the most important and least understood problems in the
construction of supersymmetric models is the mechanism of supersymmetry
breaking. 
In the context of constrained MSSM (CMSSM) it is  assumed that the soft
supersymmetry-breaking  masses $m_0$, $m_{1/2}$ and the trilinear soft
parameter $A_0$,  have universal values at the GUT scale.
One then  analyzes the impacts of the different 
phenomenological limits on the allowed values of $m_{1/2}$ and $m_0$ as 
functions of $\tan \beta$, 
assuming some default value of $A_0$ and determining the Higgs mixing 
parameter $\mu$ and the pseudoscalar Higgs mass $m_A$ by using the 
electroweak vacuum 
consistency conditions~\cite{efgos,eoss,stauco,other,more,eossgr}. 
On the other hand, specific models of 
supersymmetry breaking predict relations between these 
different soft supersymmetry-breaking parameters. For example, certain 
`no-scale' models \cite{noscale} 
may predict $m_0 = 0$ at the Planck scale.
Here we analyze (for details see Ref.~\cite{eoss}) a 
different question, namely the consistency of some proposed relations 
between $m_0$, $A_0$ and $B_0$ using for convenience
$A_0 =  {\hat A} m_0$, $B_0  =  {\hat B} m_0$.
A generic minimal supergravity (SUGRA) model \cite{sugr2} prediction is that
${\hat B} = {\hat A} -1$ \cite{mark}, and the simplest Polonyi 
model~\cite{pol} predicts 
that $\vert {\hat A} \vert = 3 - \sqrt{3}$~\cite{bfs}.

For a specific value of ${\hat A}$ and ${\hat B}$, 
these relations may be used to replace an {\it
ad hoc} assumption on the input value of $A_0$. 
For any given value of $m_{1/2}$ and $m_0$, these constraints is satisfied
for only  specific values of $\tan \beta$. Therefore, the results of
imposing these SUGRA relations  may conveniently be displayed in 
a single $(m_{1/2}, m_0)$ plane across which $\tan \beta$ varies in a 
determined manner. Furthermore, the 
phenomenological constraints on $m_{1/2}$ and $m_0$ 
can  be used to provide both upper and lower limits on the allowed 
values of $\tan \beta$~\cite{eoss}.

\begin{figure}[t]
\begin{center}
\includegraphics[height=4.5cm]{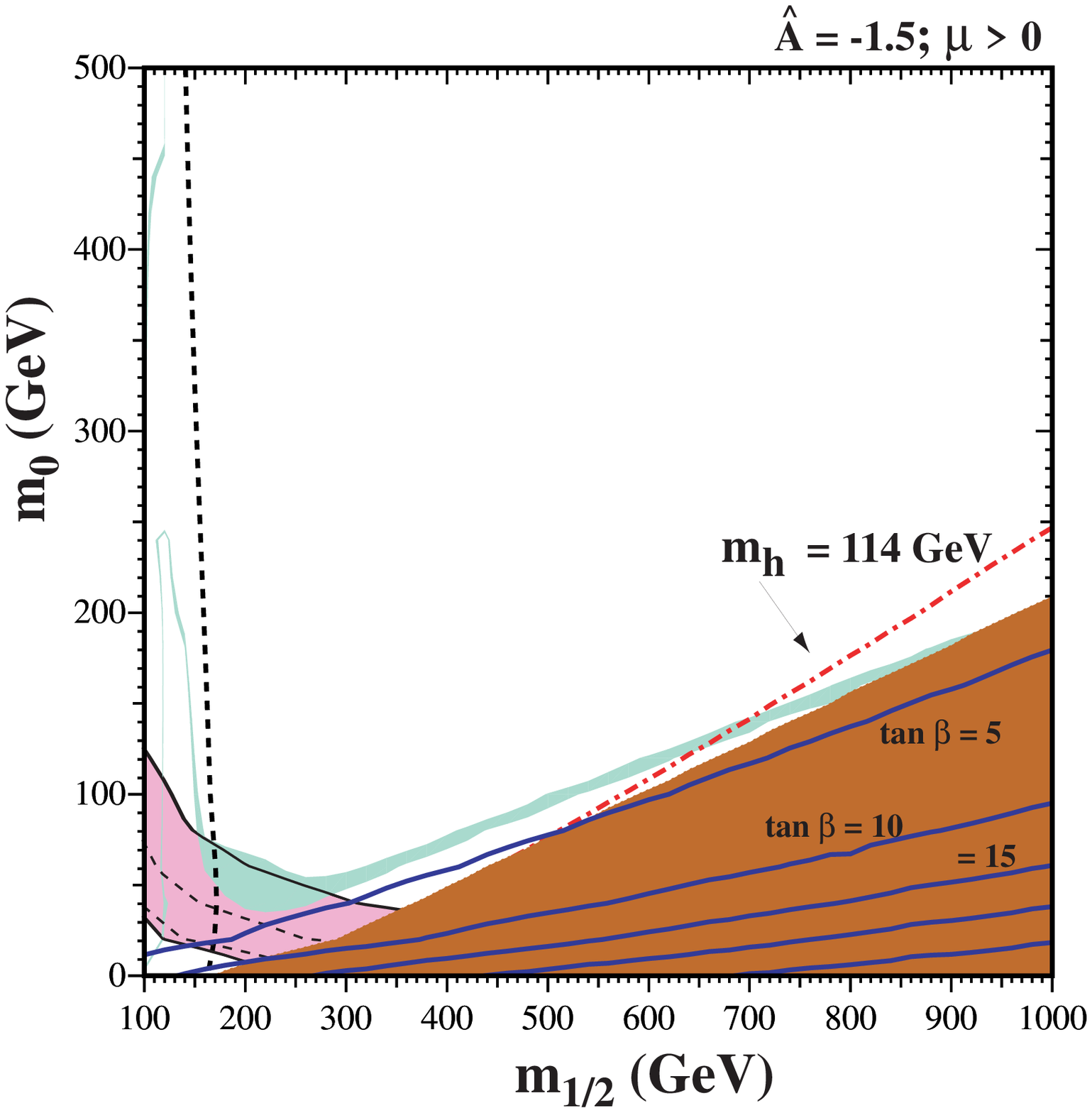}
\includegraphics[height=4.5cm]{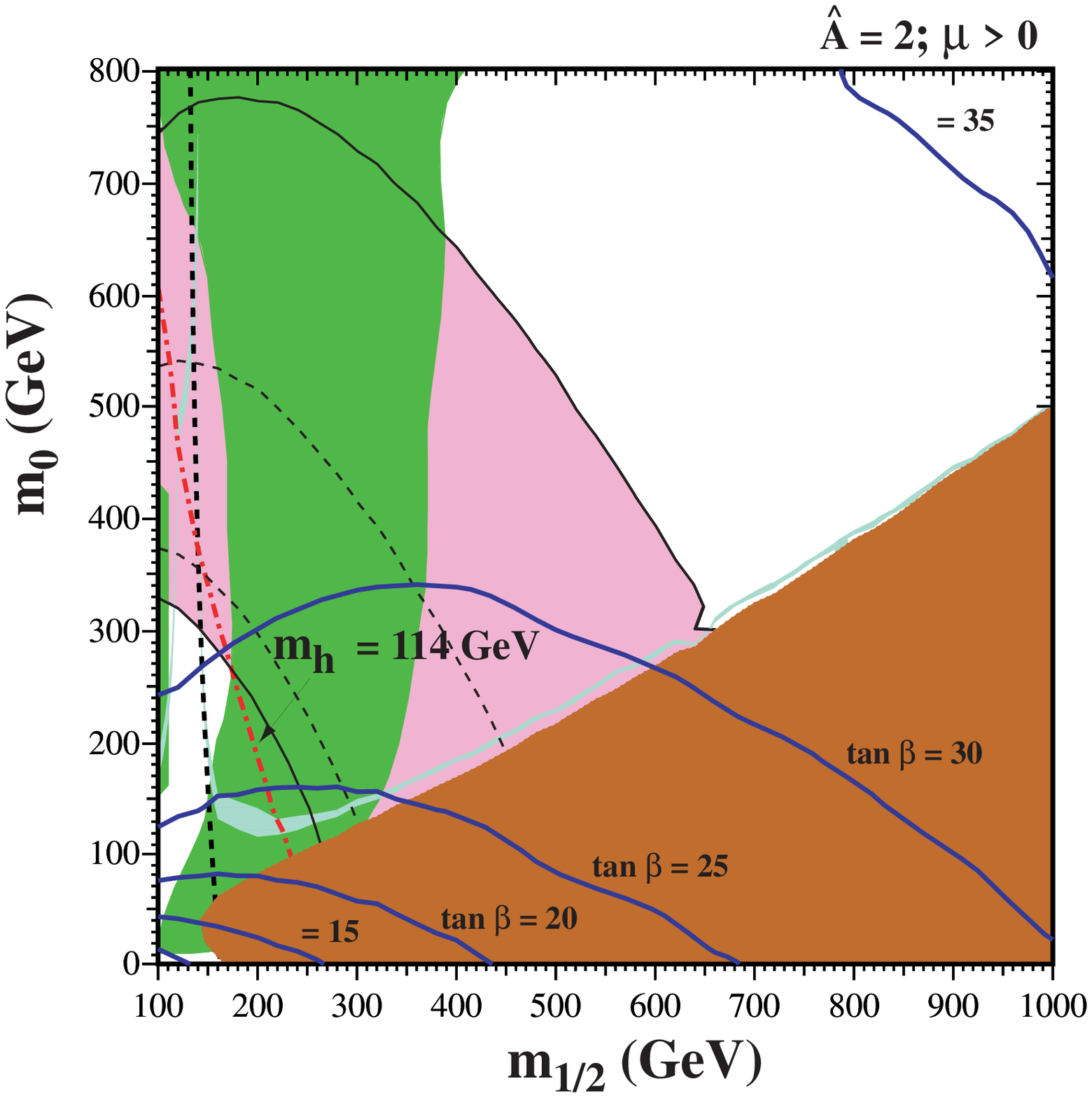}
\end{center}
\caption{\it
Examples of $(m_{1/2}, m_0)$ planes with contours of $\tan \beta$ 
superposed, for $\mu > 0$  (a) ${\hat A} = -1.5$,  
and (b) ${\hat A} = 2.0$, with ${\hat B} = {\hat A} -1$. 
In each panel, we show  regions excluded by
various constrained as they are described in the text.}
\label{fig:Polonyi}
\end{figure}

We display in Fig.~\ref{fig:Polonyi} the contours of $\tan \beta$ (solid
blue lines) in the $(m_{1/2}, m_0)$ planes for selected values of ${\hat
A}$, ${\hat B}$ and  $\mu>0$. Also shown are the contours where
$m_{\chi^\pm} > 104$~GeV (near-vertical black dashed lines) and $m_h >
114$~GeV (diagonal red dash-dotted lines). The excluded regions where
$m_\chi > m_{\tilde \tau_1}$ have dark (red) shading, those excluded by $b
\to s \gamma$ have medium (green) shading, and those where the relic
density of neutralinos lies within the WMAP range $0.094 \le \Omega_\chi
h^2 \le 0.129$ have light (turquoise) shading. Finally, the regions
favoured by $g_\mu - 2$ at the 2-$\sigma$ level are medium (pink) shaded.

As seen in panel (a) of Fig.~\ref{fig:Polonyi}, when  ${\hat
A} = -1.5$, close to its minimum possible value, the contours of $\tan
\beta$ rise diagonally from low values of $(m_{1/2}, m_0)$ to higher
values, with higher values of $\tan \beta$ having lower values of $m_0$
for a given value of $m_{1/2}$. The $m_h = 114$~GeV contour rises in a
similar way, and regions above and to the left of this contour have 
$m_h < 114$ GeV and are excluded.  Therefore,  only a very limited 
range of $\tan \beta \sim 4$ is
compatible with the $m_h$ and $\Omega_{CDM} h^2$ constraints. 
At lower values of ${\hat A}$, the slope of the Higgs contour softens and 
even less of the parameter space is allowed.  Below ${\hat A} \simeq -1.9$, 
the entire $m_{1/2} - m_0$ plane is excluded.
In panel (b) of
Fig.~\ref{fig:Polonyi}, when ${\hat A} = 2.0$, close to its maximal value
for $\mu > 0$, the $\tan \beta$ contours turn over towards smaller
$m_{1/2}$, and only relatively large values 
$25 \lesssim \tan \beta \lesssim 35$ are
allowed by the $b \to s \gamma$ and $\Omega_{CDM} h^2$ constraints,
respectively.

\begin{figure}[t]
\centering\includegraphics[height=4cm]{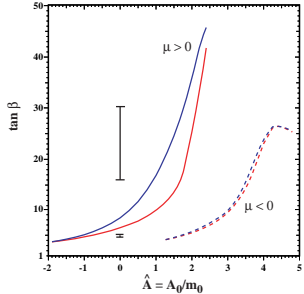}
\caption{\it
The ranges of $\tan \beta$ allowed if ${\hat B} = {\hat A} - 1$ for $\mu > 
0$ (solid lines) and $\mu < 0$ (dashed lines). The Polonyi model
corresponds to ${\hat A}  \simeq \pm 1.3$. Also shown as `error bars' 
are the ranges of $\tan \beta$ allowed in the no-scale case ${\hat A} = 
{\hat B} = 0$ for $\mu > 0$ (upper) and $\mu < 0$ (lower).}
\label{fig:tanbeta}
\end{figure}

We note the absences of both the funnel and the focus-point regions. 
In the case of the funnel, this is due to the relatively small values of
$\tan \beta$ allowed in the class of models considered here: we recall
that the funnel region appears only for large $\tan \beta \gtrsim 45$ for
$\mu > 0$ and $\tan \beta \gtrsim 30$ for $\mu < 0$ in the CMSSM.
Moreover,  as $A_0$ is increased, the focus
point is pushed up to higher values of $m_0$. Here, with $A_0 \propto m_0$, the
focus-point region recedes faster than $m_0$ if ${\hat A}$ is large
enough, and is therefore never encountered. 


It became clear that only limited ranges of $\tan 
\beta$ are consistent with the phenomenological constraints within any 
given pattern of supersymmetry breaking. We display in 
Fig.~\ref{fig:tanbeta} the ranges of $\tan \beta$ allowed as a function  
of ${\hat A}$. We find consistent solutions to 
all the phenomenological constraints only for
$- 1.9  <  {\hat A} <  2.5$,
over which range $3.7 <  \tan \beta \lesssim  46$.
In the specific case of the simplest 
Polonyi model with positive ${\hat A} = 3 - \sqrt{3}$, we find
$11 < \tan \beta  <  20$,
whereas the range in $\tan \beta$ for the negative Polonyi model 
with ${\hat A} = \sqrt{3} - 3$, is 4.4 -- 4.6.
The corresponding results for $\mu < 0$ are  $1.2 <  {\hat A}  <  4.8$
over which range $4 < \tan \beta  \lesssim  26$.
The range of ${\hat A}$ is shifted, and the range of $\tan \beta$ reduced, 
as compared to the case of $\mu > 0$. In particular, the negative Polonyi 
model is disallowed and the positive version is allowed only for $\tan 
\beta \sim 4.15$.


We have shown  that only a restricted range of $\tan \beta$
is allowed in any specific pattern of supersymmetry breaking. We have
illustrated this point by discussions of minimal SUGRA models with
${\hat A} = {\hat B} + 1$ and no-scale models with ${\hat A} = {\hat B} =
0$, but the same comment would apply to other models of supersymmetry
breaking not discussed here.

\section*{Acknowledgments}
This work was supported in part
by DOE grant DE--FG02--94ER--40823. I wish to thank my collaborators
John Ellis, Keith Olive and Yudi Santoso.


\end{document}